\documentclass[a4paper,11pt]{article}
\usepackage{amsmath,graphicx,color,mathrsfs,subfigure,multicol}

\definecolor{rosso}{cmyk}{0,1,1,0.4}
\definecolor{rossos}{cmyk}{0,1,1,0.55}
\definecolor{rossoc}{cmyk}{0,0.5,1,0.2}
\definecolor{blu}{cmyk}{1,1,0,0.3}
\definecolor{blus}{cmyk}{1,1,0,0.6}
\definecolor{blucc}{cmyk}{1,0.4,0.2,0}
\definecolor{viola}{cmyk}{0,1,0,0.6}
\definecolor{viola2}{cmyk}{0,1,0.2,0.6}
\definecolor{verde}{cmyk}{0.92,0,0.59,0.25}
\definecolor{verdec}{cmyk}{0.92,0,0.59,0.15}
\definecolor{verdes}{cmyk}{0.92,0,0.59,0.4}

\newcommand{\mSG}{m_{\tilde{G}}}

\newcommand{\eq}[1]{~{\rm (\ref{eq:#1})}}

\newcommand{\GeV}{\,{\rm GeV}}
\newcommand{\TeV}{\,{\rm TeV}}
\newcommand{\eV}{\,{\rm eV}}
\def\circa#1{\,\raise.3ex\hbox{$#1$\kern-.75em\lower1ex\hbox{$\sim$}}\,}

\newcommand{\md}[1]{\langle#1\rangle}

\newcommand{\PRL}{Phys. Rev. Lett.}

\newcommand{\PR}{Phys. Rev.}

\newcommand{\Tr}{{\rm Tr}\,}
\newcommand{\beq}{\begin{equation}}
\newcommand{\eeq}{\end{equation}}

\def\circa#1{\,\raise.3ex\hbox{$#1$\kern-.75em\lower1ex\hbox{$\sim$}}\,}
\makeatletter

\def\art{\@ifnextchar[{\eart}{\oart}}
\def\eart[#1]#2#3#4#5#6{{\rm #2}, {\em #3  #4} {\rm (#6) #5} ({\em #1})}
\def\hepart[#1]#2{{\rm #2, \em#1}}
\newcommand{\oart}[5]{{\rm #1}, {\em #2  #3} {\rm (#5) #4}}

\newcounter{alphaequation}[equation]
\def\thealphaequation{\theequation\hbox to
0.6em{\hfil\alph{alphaequation}\hfil}}
\def\eqnsystem#1{
\def\@eqnnum{{\rm (\thealphaequation)}}
\def\@@eqncr{\let\@tempa\relax \ifcase\@eqcnt \def\@tempa{& & &} \or
  \def\@tempa{& &}\or \def\@tempa{&}\fi\@tempa
  \if@eqnsw\@eqnnum\refstepcounter{alphaequation}\fi
\global\@eqnswtrue\global\@eqcnt=0\cr}
\refstepcounter{equation} \let\@currentlabel\theequation \def\@tempb{#1}
\ifx\@tempb\empty\else\label{#1}\fi
\refstepcounter{alphaequation}
\let\@currentlabel\thealphaequation
\global\@eqnswtrue\global\@eqcnt=0 \tabskip\@centering\let\\=\@eqncr
$$\halign to \displaywidth\bgroup \@eqnsel\hskip\@centering
$\displaystyle\tabskip\z@{##}$&\global\@eqcnt\@ne
\hskip2\arraycolsep\hfil${##}$\hfil& \global\@eqcnt\tw@\hskip2\arraycolsep
$\displaystyle\tabskip\z@{##}$\hfil
\tabskip\@centering&\llap{##}\tabskip\z@\cr}
\def\endeqnsystem{\@@eqncr\egroup$$\global\@ignoretrue} \makeatother

\newcommand{\SU}{{\rm SU}}

\begin{document}
\begin{flushright}
{IFUP--TH/2005-29}
\end{flushright}
\vspace{0.5cm}

\begin{center}

{\Huge \bf \color{rossos}
The Super-little Higgs}\\[1cm]

{
{\large\bf C. Cs\'aki$^a$, G. Marandella$^b$,  Y. Shirman$^{c}$ and A. Strumia$^{d}$}
}
\\[7mm]
{\it $^a$ Institute for High Energy Phenomenology \\ Newman Laboratory of Elementary Particle Physics \\ Cornell University, Ithaca, NY, 14853, USA}\\[3mm]
{\it $^b$ Department of Physics, University of California, Davis, CA 95616, USA}\\[3mm]
{\it $^c$ Theoretical Division T-8, Los Alamos National Laboratory, Los Alamos, NM 87545, USA}\\[3mm]
{\it $^d$ Dipartimento di Fisica dell'Universit{\`a} di Pisa
and INFN,
Italia}\\[1cm]
\vspace{-0.3cm}

{\tt  csaki@lepp.cornell.edu, maran@physics.ucdavis.edu,\\
shirman@lanl.gov, Alessandro.Strumia@df.unipi.it}

\vspace{1cm}

{\large\bf\color{blus} Abstract}

\end{center}
\begin{quote}
{\large\noindent\color{blus} Supersymmetry combined with
little-Higgs can render the Higgs vev super-little, providing models
of electroweak symmetry breaking free from fine-tunings. We discuss
the difficulties that arise in implementing this idea and propose
one simple successful model. Thanks to appropriately chosen Higgs
representations, $D$-terms give a negligible tree-level mass term to
the Goldstone. The fermion representations are anomaly free,
generation independent and embeddable into an SU(6) GUT. A simple
mechanism provides the large top quark mass.}
\end{quote}

\section{Introduction}
Weak-scale supersymmetry (SUSY) is considered the most promising
interpretation of the origin of the  electroweak symmetry breaking
scale. However, the non-discovery of superpartners and the Higgs at
LEP and the Tevatron imply that almost all superpartners must be
heavier than the $W,Z$ vectors, making typical SUSY models
fine-tuned~\cite{SUSYFT}. The essential problem is that the mass
parameters in the Higgs potential are determined by the soft SUSY
breaking terms. It is then hard to understand how the Higgs VEV and
consequently the $W,Z$ masses could naturally be sufficiently
smaller than the soft breaking masses themselves. This problem is
exacerbated in the Minimal Supersymmetric Standard Model (MSSM)
 by the
fact that one needs significant one loop corrections to the higgs
quartic self coupling in order to push the higgs mass above the 115
GeV LEP bound \cite{mhexp}. This can be achieved with heavy and maximally-mixed stops,
at the price of more fine-tuning.

Faced with this problem, alternative interpretations have been sought after.
One prominent idea~\cite{GK} is that the Higgs
could be a pseudo-Goldstone boson of a global symmetry broken at a scale $f$.
The main problem with this approach is that the Higgs does not in fact behave
very much like a Goldstone-boson: it has a sizable ${\cal O}(1)$ quartic self coupling, implying (naively)
that any global symmetry should be broken by ${\cal O}(1)$ effects
to generate such a coupling. This problem has been circumvented in
little Higgs models by introducing the idea of collective breaking:
the mass of the pseudo-Goldstone can be protected from large
corrections (while still allowing for a sizable quartic) if every
coupling respects some subset of the global symmetries under
which the Goldstone transforms~\cite{LH,lh}. However, in most little-Higgs models
electroweak precision observables are affected at tree level by
the new physics introduced to implement this idea,
pushing the  symmetry breaking scale $f$ in the multi-TeV range~\cite{lhew,Marandella:2005wd}
such that fine-tuning are needed to render the Higgs vev little enough~\cite{Casas}.

\smallskip

One can see that the problems with the two approaches are
complementary to each other: supersymmetric models would not give
large electroweak precision corrections for soft breaking terms of
order a few hundred GeV, however the Higgs potential is fine tuned
due to corrections to the soft masses of order $({m_{\rm soft}}/{4\pi
})^2 \ln \Lambda/m_{
\rm soft}$, where $m_{\rm soft}$ is a typical soft breaking term
and $\Lambda$ is the cutoff scale. With
a high cutoff scale $\Lambda\sim 10^{16}-10^{19}$ GeV the logarithm is
basically compensating for the loop factor and so the absence of
fine tuning would imply $m_{
\rm soft} \sim M_Z$. In little Higgs models the
corrections to the Higgs mass are generically of the form $
({f}/{4\pi})^2 \ln f/v$, where where $v$ is the Higgs VEV. This
could give a natural Higgs potential for $f \sim$ few hundred GeV,
but typically $f$ has to be much larger.
If one were able to combine the two approaches, a natural Higgs potential
may emerge:  with $m_{
\rm soft}$ and $f$ are somewhat above current experimental bounds,
$$m_{\rm soft}\sim \hbox{few
hundred GeV}\qquad\hbox{ and }\qquad f\sim\hbox{few TeV},$$
the corrections to the Higgs mass would be given by $({m_{
\rm soft}}/{4\pi })^2 \log f/m_{
\rm soft}$.
See~\cite{SU3global,Pok,Schmaltz} for recent attempts in this direction.

We stress that in this approach the hierarchy
problem would be mainly resolved by supersymmetry, and the effect of
the pseudo-Goldstone mechanism would be to ensure that the Higgs
mass parameter relevant to electroweak symmetry breaking be a loop
factor lower than the soft breaking mass parameters (without
large logarithmic enhancement) thus naturally making the $W,Z$ bosons
lighter than all supersymmetric particles.
\medskip

When using  the lack of positive experimental results
as a guideline for exploring new directions
one should pay attention in not inventing
a medicine that is worse than the illness:
a complicated model  might look
less plausible than the MSSM fine-tuned at a few $\%$ level.

In the present case, a very similar mechanism has already been applied to SUSY
models when trying to solve the doublet-triplet splitting problem of
SUSY GUTs. The perhaps most elegant resolution of this problem is to
assume that the Higgs doublets are much lighter than
the triplets because they are pseudo-Goldstone bosons. An
appealing model of this sort, based on an SU(6) gauge
group, has been suggested in~\cite{SU6,SU620} improving on~\cite{SU6glob}.
The idea is that the Higgs fields of the
model are comprised of two sectors (an SU(6) adjoint $\Sigma$ and
a pair of fundamentals $H,\bar{H}$) which are not directly connected
by superpotential terms. Then the Higgs sector has an $\SU(6)_\Sigma
\otimes \SU(6)_{H,\bar{H}}$ global symmetry which is broken via the Higgs VEV's
around the GUT scale as follows: $\SU(6)_\Sigma \to \SU(4)\otimes
\SU(2)_L\otimes {\rm U}(1)_Y$ and $\SU(6)_{H,\bar{H}}\to \SU(5)$.
One key feature of the model is that
this symmetry-breaking pattern  leaves one complex $\SU(2)_L$ doublet as Goldstone bosons (and
supersymmetry protects a second doublet from a large mass). The
other key feature of this model is that while gauging $\SU(6)$
explicitly breaks the global symmetry, due to the
non-renormalization theorem the Goldstones can not get a supersymmetric mass from
the $D$-terms and pick up a mass only after supersymmetry is softly broken.

The description of the SU(6) GUT model eerily reminds the
story of the `simplest' little Higgs model~\cite{Schmaltz}, where an
$\SU(3)_L \otimes {\rm U}(1)_X$ gauge symmetry is broken by two separate
non-communicating sectors leaving two doublets as Goldstones. Since
$\SU(3)_c\otimes \SU(3)_L\otimes {\rm U}(1)_X \subset \SU(6)$ it is most
natural to try to apply the particular group theory suggested by the
SU(6) model to our attempt at combining supersymmetry with a
pseudo-Goldstone mechanism. In fact, it turns out that this approach
has some advantages compared to the `simplest' little Higgs:
\begin{itemize}
\item Unifying two Higgs doublets in an $\SU(3)_L$ adjoint ensures that,
in appropriate ranges of parameter space, $D$-terms do not give dangerous
contributions to the  Goldstone potential.
\item The chiral fermions are automatically anomaly free,
generation independent, and embeddable into an SU(6) GUT group.
\item A simple mechanism makes the top quark
the only SM fermion with a mass comparable to the electroweak scale.
\item The gauge couplings may possibly unify.
One can find a simple matter content such that the gauge couplings unify
at a high scale around $M_{\rm Pl}$. However,
in the full model, the QCD coupling hits a Landau pole around $10^{13}\GeV$.
\end{itemize}

This paper is organized as follows. In Section 2 we give a more
quantitative description of the fine tuning problems facing
supersymmetric and little Higgs models, and a brief summary of our
approach to combining supersymmetry with the pseudo-Goldstone
mechanism. In Section 3 we present the concrete model we
investigate: we describe the gauge, Higgs and matter content, and
introduce a top sector, that provides  an order one Yukawa coupling
only to the top quark. We compute the electroweak precision bounds
on the $\SU(3)_L$ breaking scales. In Section 4 we calculate the
Higgs potential induced by one-loop effects and show how one could
generate a tree-level quartic self-interaction term for the Higgs
using a version of the sliding singlet mechanism suggested
in~\cite{Schmaltz}. In Section 5 we discuss issues related to the
phenomenology of the model: we calculate the running of the gauge
couplings and comment on possibilities for unification, we discuss
the consequences of the presence of extra sterile neutrinos, and
comment on possible signatures  at the LHC. Finally we conclude in
Section 6.

\section{Motivation}

\subsection{The problems of SUSY}
Most SUSY models are fine tuned.
In order to motivate our extensions of SUSY models it is useful to rephrase
this problem in the following way.
Let us define $f$ as the RGE energy scale at which EWSB appears:
$m^2_h (f) = 0$, where $m_h^2$ is the running Higgs mass parameter.
RGE corrections will drive $m^2_h$ negative, and in order to
 have correct EWSB they should be not too large.
In models like the constrained minimal supersymmetric standard model
(CMSSM), the scale $f$ is dynamically determined by RGE running.
Keeping only the dominant top Yukawa contribution to the running of
$m_h^2$, and assuming large $\tan\beta$ (at tree level this
corresponds to a Higgs with mass equal to $M_Z$) one needs: \beq
M_Z^2 = \frac{3y_t^2}{2\pi^2}m_{\tilde{t}}^2 \ln
\frac{f}{m_{\tilde{t}}}\eeq where $y_t = m_t/174\GeV\approx 1$ and
$m_{\tilde{t}}$ is the average stop mass. This contribution to
$M_Z^2$ would generically
 be much larger than $M_Z^2$, and it can be reduced only at the price
of fine-tuned cancellations with other contributions. To avoid such
fine tunings one needs to impose that this contribution is not
unnaturally large. To achieve this  one would need \beq
m_{\tilde{t}}\circa{<}\hbox{few hundred GeV}\qquad\hbox{and}\qquad
f\circa{<}\hbox{few TeV}.\eeq Models like the CMSSM are problematic
because the scale $f$ is dynamically determined by RGE running as $f
= M_{\rm GUT} e^{-F}$ where the function $F$ depends on all
dimensionless parameters (like $g_3,\lambda_t$, $M_{1/2}/m_{ \rm
soft}$). Then $f\sim \TeV$ can usually only be obtained thanks to a
fine-tuning among these parameters. The main motivation of this
paper is to try to find a rationale for this apparent coincidence
 of $f$ with the electroweak scale.
The simplest possible interpretation is that
the Higgs is a pseudo-Goldstone-boson of a global symmetry
spontaneously broken at a low energy scale $f$.
Models of this type with $f\sim M_{\rm GUT}$ have been studied as solutions to
the doublet/triplet problem of SUSY GUTs: this would explain why
$m_h \ll M_{\rm GUT}$, but would not make the low-energy Higgs potential
more natural.
However, recently proposed little-Higgs techniques allow us
to lower $f$ down to the TeV scale and thus obtain natural EWSB. This is the
direction we will be pursuing in this paper.
\footnote{See~\cite{lowQ, lowQ2} for alternative tentative interpretations of the smallness of $f$.}

\subsection{The problems of Little Higgs}

The Higgs boson does not seem to be the pseudo-Goldstone boson
of a global symmetry spontaneously broken at a scale $f$.
Indeed the Goldstone mechanism completely resets the Higgs potential:
 $V(h) = 0\cdot |h|^2 + 0 \cdot |h|^4$,
while the Higgs must have a substantial quartic coupling.
This problem has been recently circumvented by inventing
very specific  breakings of the global symmetry,
that generate a $|h|^4$ term at tree level,
while  $|h|^2$ only arises at loop level.
In models of this kind the generic outcome of pseudo-Goldstone models
\beq V(h)\propto \cos^n(|h|/f)\qquad\hbox{ with a minimum at
$\langle |h|\rangle\sim f$}\eeq
gets replaced by
$\langle |h|\rangle\sim f y_t/4\pi$.
In the `little-Higgs' limit $\langle |h| \rangle \ll f$ the Goldstone potential can be expanded as
\beq V(h)=- m^2 |h|^2 + \lambda |h|^4 + {\cal O}(\frac{|h|^6}{f^2}).\eeq
The mass term is generated by one-loop corrections as \beq
\label{eq:mm} m^2 = 0 \cdot \Lambda ^2 +  \frac{{\cal
O}(g^2,\lambda_t^2)}{(4\pi )^2} f^2 , \eeq
where the quadratically
divergent piece cancels thanks to the little Higgs mechanism,
leaving a logarithmically divergent contribution.
Most models assume or need a low cut-off, $\Lambda \circa{<}4\pi f$.

\medskip

However, the new physics introduced to implement the
little-Higgs mechanism, providing the
cancellation of the UV-divergent $\Lambda^2$ term, typically also gives
tree-level corrections to precision data implying $f\circa{>}$ few
TeV. As a result a sizable fine-tuning (the typical one corresponding
to the `little hierarchy problem'~\cite{LEPparadox})
is needed to render $m^2$ small enough. This situation is so bad because the actual gain achieved by
the little-Higgs mechanism is more modest than what is suggested by
the na\"{\i}ve estimate $\md{|h|}/f\sim y_t/4\pi$: the top loop happens
to be multiplied by big ${\cal O}(1)$ coefficients. Indeed without a
tree level quartic one would get $\md{|h|}/f\sim 1$, and the top loop
correction to the quartic is almost enough to get the desired higgs
mass, $m_h\circa{>} 115\GeV$~\cite{mhexp}: adding a tree level
quartic does not make the higgs as little as desired, $\md{|h|}\ll f$.
One needs to make $m^2$ smaller than its natural value given by
Eq.\eq{mm}, by invoking fine-tuned cancellation between its
contributions.

If we were to combine supersymmetry with the little Higgs mechanism
then SUSY will naturally lower $m^2$, giving \beq \label{eq:msusy}
m^2 \sim \frac{{\cal O}(g^2,\lambda_t^2)}{(4\pi )^2} m_{\rm soft}^2
\ln\frac{f}{m_{\rm soft}} .\eeq A super-little (slittle) higgs
$\md{|h|}\ll f$ is now obtained for $m_{\rm soft}\ll f$.

\subsection{SUSY and little-Higgs: a difficult marriage}

We discussed
why supersymmetric little-Higgs models are an
interesting possibility. We now discuss the obstacles that must be
overcome to actually implement this idea. The main incompatibility
is that supersymmetry suppresses $\lambda$ more than $m^2$,
providing a natural mechanism to do exactly the opposite of what the
little-Higgs mechanism is supposed to do. More concretely:
\begin{itemize}

\item  Supersymmetric $D$-terms break the global symmetry already at tree level,
making the soft pseudo-Goldstone mass term not smaller than
soft terms of other sparticles.

\item Supersymmetry forbids superpotential contributions to the quartic higgs coupling,
if the low energy theory has the MSSM field content.
\end{itemize}

The second problem can be overcome by building little-Higgs models
that provide at tree level a NMSSM-like $SH\bar H$ coupling, where
$S$ is a singlet.
The most pressing
problem is the first one. One can hope that a simple solution exists, because
 $D$-terms are not a generic breaking of the global symmetry,
 but they have a very specific structure.
Indeed  the reason why $D$-terms are present in supersymmetric
theories is the following: when a gauge symmetry is broken,
$D$-terms give mass to the scalar superpartner of the Goldstone equal to the vector
boson mass, thereby completing a SUSY massive vector multiplet.
Therefore it is worth investigating if the specific structure of the
$D$-terms can be used to bypass the first problem.

Let us start looking at the `simplest' model~\cite{Pok,Schmaltz}. The basic idea
consists in promoting $\SU(2)_L\otimes{\rm U}(1)_Y$  to
$\SU(3)_L\otimes{\rm U}(1)_X$, and breaking $\SU(3)_L$ by two pairs
of Higgs superfield triplets, $(H_1, \bar H_1)$ and $(H_2,\bar H_2)$. In the
absence of couplings between the two pairs, the potential has an
$\SU(3)_{H_1\bar{H}_1}\otimes\SU(3)_{H_2\bar H_2}$ symmetry. Since it
is partly gauged, one of the two Goldstone bosons is eaten by the
$\SU(3)_L/\SU(2)_L$ vector bosons, while the orthogonal combination
remains a pseudo-Goldstone boson. Such pseduo-Goldstone field $G$ is contained in the triplets as
\begin{align}
H_1 & = e^{i \Pi F_2 /F_1 F} (0,0,f_1/\sqrt{2}),\qquad \bar H_1 = (0,0,\bar f_1/\sqrt{2}) e^{-i \Pi F_2 /F_1 F} \\
H_2 & = e^{i \Pi F_1 /F_2 F} (0,0,f_2/\sqrt{2}),\qquad \bar H_2 = (0,0,\bar f_2/\sqrt{2}) e^{-i \Pi F_1 /F_2 F}
\end{align}
where $F_i^2 =(f_i^2+\bar f_i^2)/2, \; F^2=F_1^2+F_2^2$ and $\Pi$ is the Goldstone $3 \times 3$ matrix
\begin{equation*}
  \Pi = \left(
    \begin{array}{ll}
      0_2 & G \\
      G^\dagger & 0
    \end{array}
  \right).
\end{equation*}

The part of the $D$-term potential that  breaks the
$\SU(3)_{H_1\bar H_1}\otimes\SU(3)_{H_2\bar H_2}$ global symmetry is: \beq  V_D=
\frac{g^2}{8}\left( |H_1^\dagger \cdot H_2|^2 - |\bar{H}_1\cdot
H_2|^2 - |\bar{H}_2\cdot H_1 |^2+ |\bar H_2^\dagger \cdot \bar
H_1|^2 \right)+\cdots \eeq and its contribution to the Goldstone potential
is \beq\label{eq:H1H2} V_D= \frac{g^2}{8}(f_1^2 -
\bar{f}_1^2)(f_2^2-\bar{f}_2^2) \cos^2 \left[ \frac{\sqrt{G^\dagger
G}}{F} \left( \frac{F_1}{F_2}-\frac{F_2}{F_1}\right) \right]+\cdots.\eeq
The vevs $f_i,\bar{f}_i$ have to be computed by minimizing the full
potential, such that in general they depend on $G$.\footnote{When one of the two sectors
is heavy and supersymmetric, this reproduces at low energy
the usual MSSM quartic Higgs couplings,
such that a Goldstone quartic potential arises if $\tan\beta\neq 1$.
In this model $\tan\beta$ is determined by the vev of an extra light field,
present because the global symmetry has an extra U(1) factor,
absent in the model that will be considered in this paper.}
We see that Goldstones typically get a mass comparable to other
sparticles, because $H_i$ and $\bar H_i$ typically have different
soft masses (the difference is induced e.g.\ by top Yukawa
renormalization effects) leading to VEVs that are
non-supersymmetric: $f_i^2 - \bar f_i^2 \sim m_i^2 - \bar{m}_i^2$.
A generic supersymmetric little-Higgs model does not
provide the desired super-little Higgs vev.

\medskip

On the other hand, if the VEVs were supersymmetric, $\langle
H_1\rangle=\langle \bar{H}_1\rangle$ and $\langle H_2\rangle=\langle
\bar{H}_2\rangle$ then no potential is generated along the Goldstone
direction. One way of naturally forcing the VEVs to be supersymmetric  is by
implementing a symmetry between $H_i$ and $\bar{H}_i$ that
keeps the soft terms equal: $m_{H_i}^2 = m_{\bar
H_i}^2$~\cite{Schmaltz},
 or by building an appropriate top sector
where only one Higgs couple is involved in large Yukawa couplings~\cite{Pok}.
Indeed eq.\eq{H1H2} shows that to
preserve the Goldstone flat direction it  is enough to have
supersymmetric VEVs only in one couple: either $f_1 = \bar f_1$ or
$f_2 = \bar f_2$.
This consideration suggests that it might be
sufficient to simply replace one of the two doublet/anti-doublet pairs with
an $\SU(3)_L$ adjoint $\Sigma$, that contains a doublet/anti-doublet Higgs pair:
their soft masses are automatically forced to be equal by
$\SU(3)_L$ gauge-invariance, and one expects that this is enough for avoiding
the unwanted $D$-term potential for the Goldstones.
(Replacing both doublets with adjoints would
not allow to get the top mass). Models of this type with $f\sim
M_{\rm GUT}$ have already been explored as an elegant solution to
the doublet/triplet splitting problem~\cite{SU6}, and as a source
for the top mass~\cite{SU620}. By `deconstructing' SU(6) into
$\SU(3)_c\otimes\SU(3)_L\otimes{\rm U}(1)_X$ fragments gives a model
analogous to the `simplest' supersymmetric little-Higgs~\cite{Pok,Schmaltz},
and with universal family
charges, no gauge anomalies, and a source for the top mass.

In conclusion, we are lead to study the model  described in detail in the next section.

\section{The model}

\subsection{The gauge sector}

We consider an $\SU(3)_c\otimes\SU(3)_L\otimes{\rm U}(1)_X$ supersymmetric gauge theory. The Higgs sector, as discussed in the previous section,
consists of one $\SU(3)_L$ triplet and anti-triplet pair,
$H$ and $\bar H$, and an adjoint $\Sigma$.

Unlike in previous attempts~\cite{Pok,Schmaltz,Frampton,SchmaltzSimplest,Kong}, our assignment of fermion gauge quantum
numbers,  listed in Table~\ref{tab:charges},
is anomaly free, generation universal and reproduces the SM chiral content.
To see easily that these goals are indeed achieved
we observe that our gauge group can be embedded into $\SU(6)$,
which contains one generation of SM fermions in a $15\oplus\bar{6}\oplus\bar{6}'$ representation.
Indeed, one can immediately check that it decomposes under the usual
$\SU(5)$ gauge group as $10_{15}\oplus 5_{15}\oplus\bar 5_{\bar{6}}\oplus1_{\bar{6}}\oplus
\bar 5'_{\bar{6}'} \oplus1'_{\bar{6}'}$,
so that the chiral part is the usual $10+\bar 5$.
Decomposing it into representations of our $\SU(3)_c\otimes \SU(3)_L\otimes {\rm U}(1)_X$ group
gives the charge assignment shown in the first two parts of Table~\ref{tab:charges}.

\begin{table}
\begin{center}
\begin{tabular}{|c|ccc|}
\hline
  & $\SU(3)_c$ & $\SU(3)_L$ & ${\rm U}(1)_X$  \\
\hline
$H$ &1& 3 &$+1/3$\\
$\bar{H}$&1& $\bar{3}$ &$ -1/3$ \\
$\Sigma$ &1& 8 &$0$\\  \hline
$2\times D_{1,2,3}$   & $\bar 3$ & 1 & $+1/3$ \\
 $2\times L_{1,2,3}$ & 1 & $\bar 3$ & $-1/3$\\
 $U_{1,2,3}$   & $\bar 3$ & 1 & $-2/3$ \\
$E_{1,2,3}$   & 1 &$ \bar 3$ & $+2/3$ \\
$Q_{1,2,3}$& 3 & 3 & 0 \\ \hline
$Q'$ & $\bar 3$ & 3 & $-1/3$\\
$\bar Q'$ & 3 &$\bar{3}$ & $+1/3$\\
\hline
\end{tabular}
\caption{\em Charge assignments of the chiral superfields.
The upper group is the Higgs sector, the middle group is the chiral matter,
the lower group is the vector-like top.}
\label{tab:charges}
\end{center}
\end{table}

The superpotential of the Higgs sector of our model is\footnote{We normalize the super-fields
such that their kinetic terms are $\int d^4\theta\,[2{\rm Tr}\,\Sigma^\dagger\Sigma + H^\dagger H + \bar{H}^\dagger \bar{H}$].}
\begin{eqnarray}
W_{\rm Higgs} &=&  \frac{M}{2}\Tr \Sigma^2 +\frac{\lambda}{3}\Tr \Sigma^3 + S (\lambda'' H\bar{H} - M^{\prime2}) + \lambda' \bar H \Sigma H
\end{eqnarray}
The dimensionful $M,M'$ terms are analogous to the MSSM $\mu$ term.
The $\SU(3)_L$-breaking Higgs VEVs are
\beq
\langle\Sigma\rangle = {\rm diag}\,(w/2,w/2,-w),\qquad \langle H\rangle  = (0,0,f/\sqrt{2}),\qquad\langle\bar H\rangle = (0,0,\bar f/\sqrt{2})\label{eq:VEVs}\eeq
and break the gauge group to the SM gauge group, giving the following masses to the extra
gauge bosons, a complex weak doublet $W'$ and a weak singlet $Z'$:
\beq \label{eq:W'Z'} M_{W'}^2 = \frac{g^2}{2}(F^2 + 9w^2), \qquad M_{Z'}^2 =\frac{2F^2g^2}{3- t^2}\eeq
where $F^2 = (f^2 + \bar f^2)/2$ and $t=\tan\theta_{\rm W}$.
The unbroken ${\rm U}(1)_Y$ arises as $Y=X+T_8/\sqrt{3}$ (on triplets $T_8/ \sqrt{3} = \,{\rm diag}(1/6,1/6,-1/3)$).
The hypercharge gauge coupling is $1/g^{\prime 2} = 1/g_X^2 +1/3g^2$,
and $g$ is the usual $\SU(2)_L$ coupling.

\bigskip

We will need VEVs an order of magnitude larger than the SUSY-breaking scale.
They can arise because  $M,M'$ are larger than the SUSY-breaking scale, or because
the $\lambda,\lambda''$ couplings are small.
The model can work even with a vanishing $M$.

\smallskip

For $\lambda'=0$
the superpotential decomposes as $W(H,\bar H) + W(\Sigma)$,
acquiring a $\SU(3)_{H,\bar H}\otimes\SU(3)_\Sigma$ global symmetry.
The Higgs sector consists of 4 complex doublets, out of which one is eaten
during the breaking of $\SU(3)_L$
and one acquires a heavy mass from the $D$-terms, forming a massive supersymmetric vector multiplet
with the corresponding gauge bosons.
We are then left with two complex doublets, $H_{\rm u},H_{\rm d}$ which are contained in the $\SU(3)_L$ multiplets as follows
\begin{eqnsystem}{sys:Goldstones}
  H &=& \exp \left( i \Pi \frac{3 w}{F V}  \right) \langle H \rangle, \\
  \bar H &=&   \langle \bar H \rangle \exp \left(-i \Pi \frac{3 w}{F V} \right) \\
  \Sigma &=& \exp \left(-i \Pi \frac{F}{3 w V} \right) \langle \Sigma \rangle \exp \left(i \Pi \frac{F}{3 w V} \right)
\label{eq:Sigma}
\end{eqnsystem}
where $F^2\equiv (f^2 + \bar f^2)/2,\; V^2\equiv F^2+9w^2$, the VEVs are defined in Eq.~(\ref{eq:VEVs})
and $\Pi$ is a $3 \times 3$ matrix defined as
\begin{equation}
  \label{eq:Pi}
  \Pi = \frac{1}{\sqrt{2}} \left( \begin{array}{ll} 0_2 & H_{\rm u} \\ H_{\rm d}^t & 0 \end{array} \right).
\end{equation}
Only the combination $G \equiv (H_{\rm u}+H_{\rm d}^\dagger)/\sqrt{2}$ is a Goldstone boson, and this is the field
 we want to identify the Higgs with. The orthogonal sGoldstone
  $\tilde{G} \equiv (H_{\rm u}-H_{\rm d}^\dagger)/\sqrt{2}$ is
massless as long as supersymmetry is unbroken and gets a mass of the order of the soft susy breaking mass for the Higgs.

\smallskip

The $\lambda'$ term explicitly breaks the global symmetry giving a
tree level mass to the Goldstone. Since a successful model does not
require $\lambda'$ to be  smaller than a `typical' SM Yukawa
coupling --- $\lambda' \circa{<}10^{ -2}$ is small enough --- it
seems unnecessary to  invent a mechanism that suppresses $\lambda'$.
From this point of view, the situation is simpler than in the
GUT-scale version of the model.

Gauge and top Yukawa interactions break the $\SU(3)_{H,\bar H}\otimes\SU(3)_\Sigma$ global symmetry, generating a mass and a quartic coupling for the pseudo-Goldstone $G$.

\subsection{The top sector}

As usual in supersymmetric theories we will assume the presence of
matter parity: a $Z_2$ symmetry under which ``matter'' chiral
superfields flip sign, while the fields in the Higgs sector are
even. Or we could impose $R$-parity, under which also the superspace
coordinate $\theta_\alpha$ flips sign. These will forbid $B,L$
violating operators. In little Higgs models one can try to impose a
different $Z_2$ symmetry, usually referred to as
$T$-parity~\cite{Tparity}, in order to forbid tree-level corrections
to electroweak precision observables. In our case this would
correspond to imposing every new particle beyond the SM to be odd:
this is neither necessary nor possible. It is not possible because
this symmetry is not compatible with the $\SU(3)_L\otimes {\rm
U}(1)_X$ gauge group. It is not necessary because we do not need to
lower the little-Higgs scale $F,w$ below a few TeV.

\medskip

After imposing this symmetry, the most general renormalizable superpotential
term involving the matter fields is
\begin{equation}
W_{\rm matter}=\alpha_{ij} Q_i \bar H D_j +\beta_{ij} E_i \bar H L_j.
\label{eq:mattersuppot}
\end{equation}
The VEV of $\bar H$ gives a mass term that mixes the extra set of $D,L$ fields
 with $\SU(3)_L$ partners of SM fields. Thus below the little-Higgs scale
the matter content of the theory is exactly that of the MSSM, plus possibly two sterile
neutrinos per generation from the third components of the $L$ fields. We will comment on these
sterile neutrinos later on. One can also see that at the renormalizable level (\ref{eq:mattersuppot})
does not lead to any mass terms for the MSSM fields even after $\SU(2)_L$ breaking. Perhaps the
simplest way to see this is by performing an $\SU(3)_L$ rotation on the $\bar H$ field such that its VEV (after
$\SU(2)_L$ breaking) is still contained in the third component. Then all of the $\SU(2)_L$ breaking would be
rotated into the $\Sigma $ field, which however does not participate in (\ref{eq:mattersuppot}).
Thus at this level all the MSSM fermions remain massless.

Therefore in order to implement the little-Higgs mechanism we must
find a way of obtaining a top Yukawa coupling compatible with the
gauge symmetries of our model. This can be  elegantly achieved
following the mechanism~\cite{SU620} employed in the SU(6) model: by
adding the appropriate vector-like representation listed in the
lower rows of Table~\ref{tab:charges},  additional Yukawa couplings
are allowed, such that one up-type quark gets a mass $m_t\sim v$.
These extra fields form a heavy vector-like top analogous to the one
employed by little-Higgs models, but different from the one employed
in the closely related `simplest' little Higgs.

Let us first review the details of this mechanism in the SU(6)
theory. We add the representation 20 (three index antisymmetric
tensor) to the other matter fields $3\times (15+\bar{6}+\bar{6}')$.
Then the most general renormalizable superpotential is
\begin{equation}
  \label{eq:yukawa}
  W= 20 \, \Sigma\,  20 + 20\, H \,15+  15\, \bar H\, \bar 6
\end{equation}
where here $H,\bar H,\Sigma$ denote the obvious $\SU(6)$-extensions of our $H,\bar H,\Sigma$ fields.
The 20 representation is pseudo-real: despite being non-chiral, SU(6) invariance keeps it massless.
The way the large Yukawa coupling is generated in the SU(6) model is by assuming that the VEV of $H,\bar{H}$
break the symmetry to SU(5). Under this subgroup
$$20 = 10_{20}+\overline{10}_{20},\qquad 15= 10_{15}+5_{15},\qquad
\bar{6}= \bar{5}_{\bar 6}+1_{\bar 6},\qquad
\bar{6}'= \bar{5}'_{\bar 6'}+1'_{\bar 6'}.$$
When $H,\bar H$ get a VEV, the Yukawa coupling $20\, H \,15$ gives
a  $\overline{10}_{20} 10_{20}$ mass term,
and $15\, \bar H\, \bar 6$ gives a $5_{15}\,\bar{5}_{\bar 6}$ mass term.
The light chiral fermions arise as $10_{20}\oplus\bar 5'_{\bar 6'}$.
The first term of eq.\eq{yukawa} provides
the top Yukawa coupling to the Higgs Goldstone in $\Sigma$.
This mechanism has the collective
symmetry breaking already built in.

Implementing this mechanism in the $\SU(3)_c\otimes \SU(3)_L\otimes
{\rm U}(1)_X$ model corresponds to adding the representations $Q' =
(\bar 3,3)_{-1/3}$ and $\bar Q' = (3, \bar 3)_{1/3}$ (which make up
the 20 of SU(6) up to two singlets). Therefore, in terms of the
$\SU(3)_c\otimes \SU(3)_L\otimes {\rm U}(1)_X$ fragments the top
sector is given by \beq W_{\rm top} = \lambda_1 \bar{Q}' \Sigma Q' +
\lambda_2 \bar{Q}' HU + \lambda_3 Q  H Q' \eeq The top sector
contains  the SM top quark and 3 additional heavy vector-like
particles. It has the collective breaking property that any
individual coupling respects the global $\SU(3)_\Sigma\otimes
\SU(3)_H$ symmetry, broken only by the simultaneous presence of both
$\lambda_1$ and $\lambda_2$ (or $\lambda_1$ and $\lambda_3$), thus
eliminating the quadratically divergent loop effect from the top
sector.

A  $ \bar{Q}'  Q' $ mass term is also allowed,
and should be not much bigger than $\lambda_1\md{\Sigma}$.
Its presence would modify the detailed expressions for the masses and for the loop induced
potential calculated below, however qualitatively does not change the story: the symmetry breaking is still collective.
In the SU(6) unified theory this mass term is not allowed,
and it remains vanishing if the 20 does not couple to the SU(6)-breaking Higgses.

\bigskip

The top mass matrix is given by
\begin{equation*}
   W_{\rm top} = \left(\bar Q', Q \right) M_{\rm top} \left(Q',U \right)^t\qquad
     M_{\rm top} =   \left( \begin{array}{cc} \lambda_1 \Sigma & \lambda_2 H \\ \lambda_3 \Omega & 0  \end{array} \right)
\end{equation*}
where $\Omega_{ij} = \epsilon_{ijk} H_k$ is a $3 \times 3$ matrix.
To find the spectrum we diagonalize the matrix $M_{\rm top}^\dagger M_{\rm top}$.
Neglecting $\SU(2)_L$ breaking, one gets a massless mode (to be identified with the SM top)
and three heavy modes (`heavy tops') with masses
\begin{equation}
M_{T_1}^2 = \lambda_1^2 w^2+ \frac{1}{2} \lambda_2^2 f^2, \;\;\;\; M_{T_{2,3}}^2 = \frac{1}{4} \left( \lambda_1^2 w^2+ 2 \lambda_3^2 f^2 \right).
\end{equation}
The SM top Yukawa coupling $y_t$ arises in terms of $\lambda_{1,2,3}$ as
\beq  y_t = \frac{f^2 \sqrt{F^2 + 9w^2} \lambda_1 \lambda_2 \lambda_3}{F \sqrt{2(2w^2 \lambda_1^2 + f^ 2\lambda_2^2)(w^2 \lambda_1^2 + 2f^2 \lambda_3^2)}} \simeq\left\{\begin{array}{ll}
\lambda_1/2 &\hbox{if $w\to 0$}\\
3f^2\lambda_2 \lambda_3/(2Fw\lambda_1) & \hbox{if $F\to 0$}
\end{array}\right.  ~.\eeq

At this stage only the top is massive, while all other Standard Model
fermions remain massless. Their masses can arise from non-renormalizable
superpotential operators  in a way similar to the construction of
\cite{SU620}, where fermion masses are suppressed by powers
of $M_{\rm GUT}/M_{\rm Pl}$.
In our model the natural suppression is given by $\epsilon=f/\Lambda^\prime$, where
$\Lambda^\prime$ is some scale associated with the explicit breaking of the global
$\SU(3)_{H,\bar H}\otimes \SU(3)_\Sigma$ symmetry.
We will see in Section \ref{quartic} that such a scale
is necessarily present in order to generate a tree-level quartic self coupling for the
Higgs, so it is natural to assume that the same scale will appear in these non-renormalizable
operators (most of which will break $\SU(3)_{H,\bar H}\otimes \SU(3)_\Sigma$).

\subsection{Electroweak precision constraints}

Precision data are affected at tree level in two different ways:
by the $Z'$, and by the presence of a Higgs triplet coupled to the Higgs doublet.
We can neglect the extra corrections induced at loop level by
supersymmetric particles and by the heavy vector-like tops.

The effects of the $Z'$ have already been studied in~\cite{Marandella:2005wd},
and data imply the bound $F>3 \TeV$ at 99\% C.L.
The Higgs triplet effects can be computed from eq.\eq{Sigma}:
expanding it in the limit $H_{\rm u},H_{\rm d}\ll w\ll F$ one finds that the  $\SU(2)_L$ adjoint contained in $\Sigma$
gets a VEV $\langle\Sigma\rangle^{ij}=H_{\rm d}^i H_{\rm u}^j/12 w$.
Assuming that $\SU(2)_L$ is dominantly broken by the Goldstone VEV,
such that $\md{H_{\rm u}}=\md{H_{\rm d}} = (0,v/\sqrt{2})$, one gets a correction
$\hat T = v^2/144w^2$ to the $\hat T$ parameter (normalized as in~\cite{STWY}).
Data roughly demand $\hat T\circa{<}10^{-3}$~\cite{STWY} implying $w \circa{>}0.5\TeV$.

\section{The Higgs potential and electroweak symmetry breaking}\label{Potential}

We now study if the motivations discussed above
can be realized in the present model,
making sparticles naturally heavier than present experimental bounds.

To start, one can verify that the $D$-terms potential
\beq V_D \propto V_D(H,\bar H) + V_D(\Sigma)  -  \bar{H}[\Sigma,\Sigma^\dagger]\bar{H}^\dagger +H^\dagger [\Sigma,\Sigma^\dagger]H\eeq
despite breaking at tree level the $\SU(3)_{H,\bar H}\otimes\SU(3)_\Sigma$ global symmetry,
does not give any contribution to the Goldstone potential:
$|G|^2$ does not appear in $[\Sigma,\Sigma^\dagger]$. However, the full potential
have to be minimized with respect to heavy fields: in the supersymmetric limit
this eliminates the $D$-terms corresponding to the broken generators.
For example, in the limit $f\gg w, m_H,m_{\rm H}$, the low-energy theory
below the scales $f,\bar{f}$ only contains the $\SU(2)\otimes {\rm U}(1)$ $D$-terms,
plus soft terms of order $(m_H^2-m_{\bar{H}}^2) D_8$ \cite{Hitoshi,Pok},
where $m_H,m_{\bar{H}}$ are the soft breaking terms for $H,\bar{H}$, 
and $D_8$ denotes the $D$-term corresponding to the $T_8$ generator of $\SU(3)$ 
with only the light fields included.
This produces a negligibly small Goldstone mass,  of order
\begin{equation}
\frac{(m_H^2-m_{\bar{H}}^2)(f^2-\bar{f}^2)w^2}{(f^2+\bar{f}^2)^2} |G|^2.
\end{equation}

The fact that no large mass for the Goldstone is
generated by the $D$-terms is a necessary but not a sufficient condition,
because it does not guarantee that the Goldstone Higgs doublet plays
the main r\^ole in electroweak symmetry breaking.
The main r\^ole would be played by non-Goldstone Higgs fields,
if their squared mass parameters are negative
(or more precisely if the full squared Higgs mass matrix has a negative eigenvalue).

Let us therefore study the full Higgs potential.
At energies below $F,w$ the little-Higgs sector decouples leaving the
MSSM with specific values of its Higgs parameters.
At tree level and for $\lambda'=0$ there is a flat Goldstone
direction: $H_{\rm u} = H_{\rm d}$ i.e.\ $\tan\beta=1$.
Therefore the Higgs mass matrix, in the basis $(H_{\rm u},H_{\rm d}^\dagger)$ has the form~\cite{Rattazzi}
\beq\label{eq:massm} \begin{pmatrix} m_{h_{\rm u}}^2 + \mu^2 & -B\mu \cr -B\mu & m_{h_{\rm d}}^2 + \mu^2 \end{pmatrix} = \frac{1}{2}
\begin{pmatrix} \mSG^2 + m_{G\tilde{G}}^2 & -\mSG^2 \cr -\mSG^2 & \mSG^2-m_{G\tilde{G}}^2 \end{pmatrix}\eeq
where $\mSG^2$ and $m_{G\tilde{G}}^2$ are given in terms of high-energy parameters as:
\beq  \mSG^2 = \frac{F^2}{V^2} m_\Sigma^2 + \frac{9w^2}{F^2 V^2} (m_H^2 f^2+m_{\bar H}^2\bar f^2)
,\qquad m_{G\tilde{G}}^2 \sim m_H^2 - m_{\bar H}^2. \eeq where
$m_\Sigma,m_H,m_{\bar H}$ denote soft mass terms. We only showed the
most important feature of the full expression of $m_{G\tilde{G}}^2$. The
Goldstone/sGoldstone mixing  $m_{G\tilde{G}}^2$ tends to induce a VEV along
the $\tan\beta\neq 1$ direction.
In this way a large enough Higgs mass can be achieved, but the
motivation for the model is lost: the Higgs is no longer a
pseudo-Goldstone boson and the low energy theory is neither better
nor worse than the generic MSSM.

This problem does not arise
in a well defined region of the parameter space.
Choosing $\mSG\sim m_\Sigma \sim \TeV$ a few times larger than $m_{H},m_{\bar{H}}$
suppresses the unwanted negative eigenvalue $-m_{G\tilde{G}}^4/4\mSG^2$ of eq.\eq{mm}.
Furthermore, $m_{G\tilde{G}}^2$ can be naturally small:
with $w$ a few times smaller than $f$,
the top Yukawa coupling is dominantly generated by $\lambda_1$:
$\lambda_2,\lambda_3$ can be small enough such that RGE running
does not break an initial equality  $m_H^2=m_{\bar H}^2$.
The initial equality naturally arises in models of soft terms
(like gauge mediation and possibly gravity mediation)
where the source of supersymmetry breaking couples equally to $H$ and $\bar H$.
Furthermore, we  assume $\mSG^2>0$.
This is a plausible assumption, since our model does not need many Yukawa couplings as large as the top Yukawa, that renormalize soft squared masses making them negative.

\bigskip

\subsection{The top loop effect}
At this point we can turn on the loop effects that break
the global symmetry, and study if Goldstones acquire a small $\SU(2)_L$-breaking VEV.
One loop corrections renormalize the parameters in eq.\eq{massm}
and more importantly generate
a potential for the Goldstone mode.
Expanding the potential along the Goldstone direction
as $V(G) = -m^2 |G|^2 + \lambda |G|^4$, it receives at one loop order two main contributions:
from the sector that provides the top Yukawa, and from gauge interactions.
The first contribution is
\begin{eqnsystem}{sys:toploop}
\lambda &\simeq & \frac{3y_t^4}{8\pi^2}(\ln\frac{m_{\tilde{t}}}{m_t} +\frac{A_t^2}{2m_{\tilde{t}}^2}(1-
\frac{A_t^2}{12m_{\tilde{t}}^2}))\\
 m^2 &\simeq& \frac{3}{8\pi^2}y_t^2 (m_{\tilde t}^2-m_t^2) (2 \ln \frac{M_{T_2}}{m_{\tilde t}} + 1 + 2c\ln\frac{M_{T_1}}{M_{T_2}})
 \end{eqnsystem}
where we assumed a common stop mass $m_{\tilde{t}}$, $A_t$ is the top $A$-term, and
\beq c=\frac{f^2 \left(F^2+9 w^2\right) \lambda _1^2 \left(2 w^2 \left(\lambda _2^2-\lambda _3^2\right) \lambda _1^2+f^2 \lambda
   _2^2 \left(\lambda _2^2-4 \lambda _3^2\right)\right)}{6 F^2 \left(2 w^2 \lambda _1^2+f^2 \lambda _2^2\right) \left(2
   \left(\lambda _2^2-\lambda _3^2\right) f^2+3 w^2 \lambda _1^2\right) y_t^2}
\eeq
As expected  $\lambda$ depends only on MSSM physics and takes its standard value,
and $m^2$ depends
only on physics up to ${\cal O}(M_{T_{1,2}})$ energies:
it is generated by one-loop
threshold corrections at the scale $M_{T_{1,2}}$, and by short RGE running
down to $m_t$, in agreement with the estimate in eq.\eq{msusy}.

Adding also the gauge contribution and making $m^2$ a few times
smaller by appropriately chosen gaugino masses, the Goldstone
potential can develop a minimum. However, even in the most favorable
situation of maximal stop mixing $A_t = \sqrt{6}m_{\tilde{t}}$,
 to get $m_h\circa{>}115\GeV$ one needs $\tan\beta\circa{>}3$
 i.e.\ the model is allowed only when its motivation is lost.

\subsection{A tree level quartic\label{quartic}}

To reach the goal outlined in the introduction, we
must introduce a little-Higgs like mechanism that provides an extra tree-level
contribution to the quartic potential.
This can be done analogously to~\cite{Schmaltz}.

We assume that the low energy theory contain extra singlet(s) $S$, such that
a NMSSM-like $SG^2$ superpotential term is allowed.
However, it is not easy to obtain only this term from the original theory.
For example  breaking the $\SU(3)_{H,\bar H}\otimes\SU(3)_\Sigma$ global symmetry by
adding terms like $S \bar H{\Sigma} H$ would also generate a Goldstone mass term of order $F,w$.
Indeed, assuming for simplicity $\bar f=f$,
\begin{equation}
\bar{H}\Sigma H = -\frac{f^2 w}{2} +\frac{V^2}{12w}   |G|^2
\end{equation}
and so $|\bar{H}\Sigma H|^2$  contains an unwanted mass term of order $f^2,w^2$ for the Goldstone.

One needs operators that contain only the Goldstone fields, without
the constant  VEVs $f,w$. One can eliminate these VEVs by adding
extra singlets, that ``slide'' absorbing the $\Sigma,H\bar H$ VEVs,
analogously to the sliding singlet mechanism~\cite{sliding} for the
doublet-triplet splitting problem of SUSY GUTs. Naively, this could
be achieved by an operator of the form $S(S'+ \bar{H}\Sigma H)$. The
$S$ equation of motion fixes  $\langle S' \rangle = -\langle
\bar{H}\Sigma H \rangle$, eliminating the $\Sigma,H,\bar H$ VEVs, and
thus leaving the Goldstone massless. However  $S'$ eliminates also
the full dependence on the Goldstone, and thus no potential is
generated for $G$. To be successful, we need a ``collective
version'' of this sliding singlet mechanism, whereby only the VEVs
get absorbed, while Goldstone interactions survive. One successful
operator is \beq\frac{1}{\Lambda} (S H + S' \frac{\Sigma
H}{\Lambda'})(\bar S \bar H + \Sigma \bar H).\label{eq:opera}\eeq
Indeed the $S$ and $S'$ equations of motion are respectively
\begin{eqnarray}
&& \bar S \bar H H +\bar{H}\Sigma H =0, \nonumber \\
&& \bar S \bar{H}\Sigma H +\bar{H}\Sigma^2 H =0.
\end{eqnarray}
Expanding these equations at leading order in the Goldstone field gives
\begin{eqnarray}
&& \bar S f^2 +(- w f^2+V^2 |G|^2/6w) =0, \nonumber \\
&& \bar S (- w f^2 +V^2 |G|^2/6w) + (f^2 w^2 -V^2 |G|^2/6) =0.
\end{eqnarray}
The Goldstone appears differently in the two equations,
while the VEVs appear in the same way:
 $\bar{S}$ `slides' to $\langle \bar S\rangle = w$ eliminating only the VEVs,
 such that the resulting potential does not contain the unwanted $|G|^2$ mass term,
 but contains the desired $|G|^4$ coupling with  ${\cal O}(1)$ coefficient.
The peculiar operator in eq.\eq{opera} could be  naturally generated
by tree-level exchange of an extra heavy pair of $\SU(3)_L$
triplets.

\medskip

Another simple mechanism for obtaining a tree-level quartic has been
suggested in~\cite{Pok}, that exploited the fact that, 
if one of the Higgs sectors have large VEVs and universal soft breaking terms,
the effective theory below the Higgs VEV only contains the $D$-terms
of the unbroken gauge generators, but with $\SU(3)$-symmetric soft-terms.
Since the MSSM $D$-terms do not give a mass for the Goldstone, this offers
the possibility of using the usual MSSM quartic for $\tan \beta >1$.
In our model this would correspond to considering the limit $w\gg
f$, such that the effective theory below $w$ contains the
$\SU(2)_L\otimes {\rm U}(1)_X\otimes {\rm U}(1)_{8}$ $D$-terms (the
broken $D$-terms are eliminated in the process of integrating out the heavy fields). 
However, the $\SU(2)_L\otimes
{\rm U}(1)_X\otimes {\rm U}(1)_{8}$ $D$-terms give the potential
\begin{equation}
-\frac{g^2 (f^2-\bar{f}^2)^2}{4 (f^2+\bar{f}^2)} |G|^2 +
\frac{2 g^2 (f^2-\bar{f}^2)^2}{3 (f^2+\bar{f}^2)^2} |G|^4
\end{equation}
for the Goldstone $G$. Thus the quartic is sizeable only if
$f$ and $\bar{f}$ deviate ${\cal O}(1)$ from each other, but in this
case a large Goldstone mass is also introduced.
To implement this mechanism in our model, one
would have to introduce another pair of triplets
$\tilde{H},\bar{\tilde{H}}$ with large VEV's and ${\cal O}(1)$
couplings to $\Sigma$ (in order to avoid the appearance of another
global $\SU(3)$ symmetry). 
With these additional fields the low-energy effective theory only contains the
$\SU(2)\otimes {\rm U}(1)$ $D$-terms, that provide only the
quartic (but no mass term) for the Goldstone
\begin{equation}
\frac{g^2 (f^2-\bar{f}^2)^2}{8 (f^2+\bar{f}^2)^2} |G|^4
\end{equation}
which is the usual MSSM quartic at $\tan \beta >1$. The value of $\tan
\beta $ is given by the ratio $f/\bar{f}$, which is mainly determined
by the soft breaking terms for $H,\bar{H}$, and in this limit can be made
different from one without introducing large corrections to the
Goldstone mass.
In the limit $w\gg f,\bar f$, 
by choosing an appropriately small $\lambda_1$ and appropriately large
$\lambda_2,\lambda_3$ our top sector provides a large top Yukawa coupling
$ y_t \le 3f \lambda_2\lambda_3/\sqrt{2}F(\lambda_2+2\lambda_3)$.


%


\section{Phenomenology}

\subsection{Gauge coupling unification?}
We now turn to the question of the unification of the gauge
couplings. We recall that in order to achieve an anomaly free matter
content we have chosen to embed the gauge group of our model
$\SU(3)_c\otimes \SU(3)_L\otimes {\rm U}(1)_X$ into a GUT $\SU(6)$
gauge group with the Standard Model fermions transforming in the
$15\oplus\bar{6}\oplus\bar{6}'$ representation. Furthermore, we have
added an extra chiral fields in the representation $20$ of $\SU(6)$
in order to obtain the large top quark mass compatibly with the
little Higgs mechanism. It is therefore natural to ask whether
unification of couplings could be achieved with an appropriate
embedding of the Higgs sector into $\SU(6)$. In order to find
unification, we can complete the $H$ and $\bar H$ multiplet into $6$
and $\bar 6$ representations of $\SU(6)$ and assume that the color
triplets contained in these fields obtain their mass at a low scale.
Finally, $\Sigma$ can be embedded into the adjoint of the $\SU(6)$
gauge group: we will assume that all additional components of
$\Sigma$ have large masses --- this constitutes an analog of the
doublet-triplet splitting problem.

\begin{figure}[t]
\begin{center}
\subfigure[]{\includegraphics[width=0.49\textwidth]{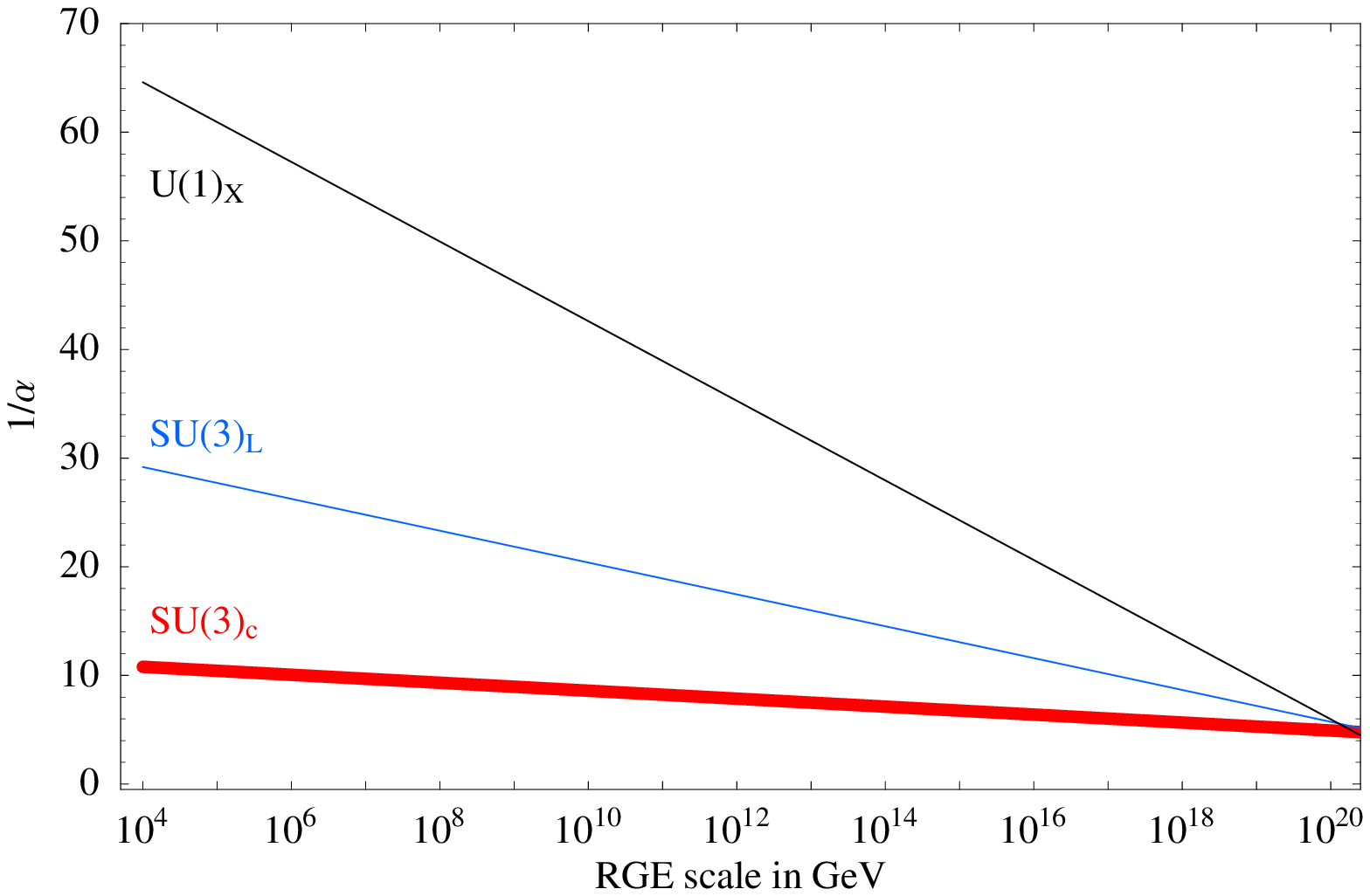}}
\subfigure[]{\includegraphics[width=0.49\textwidth]{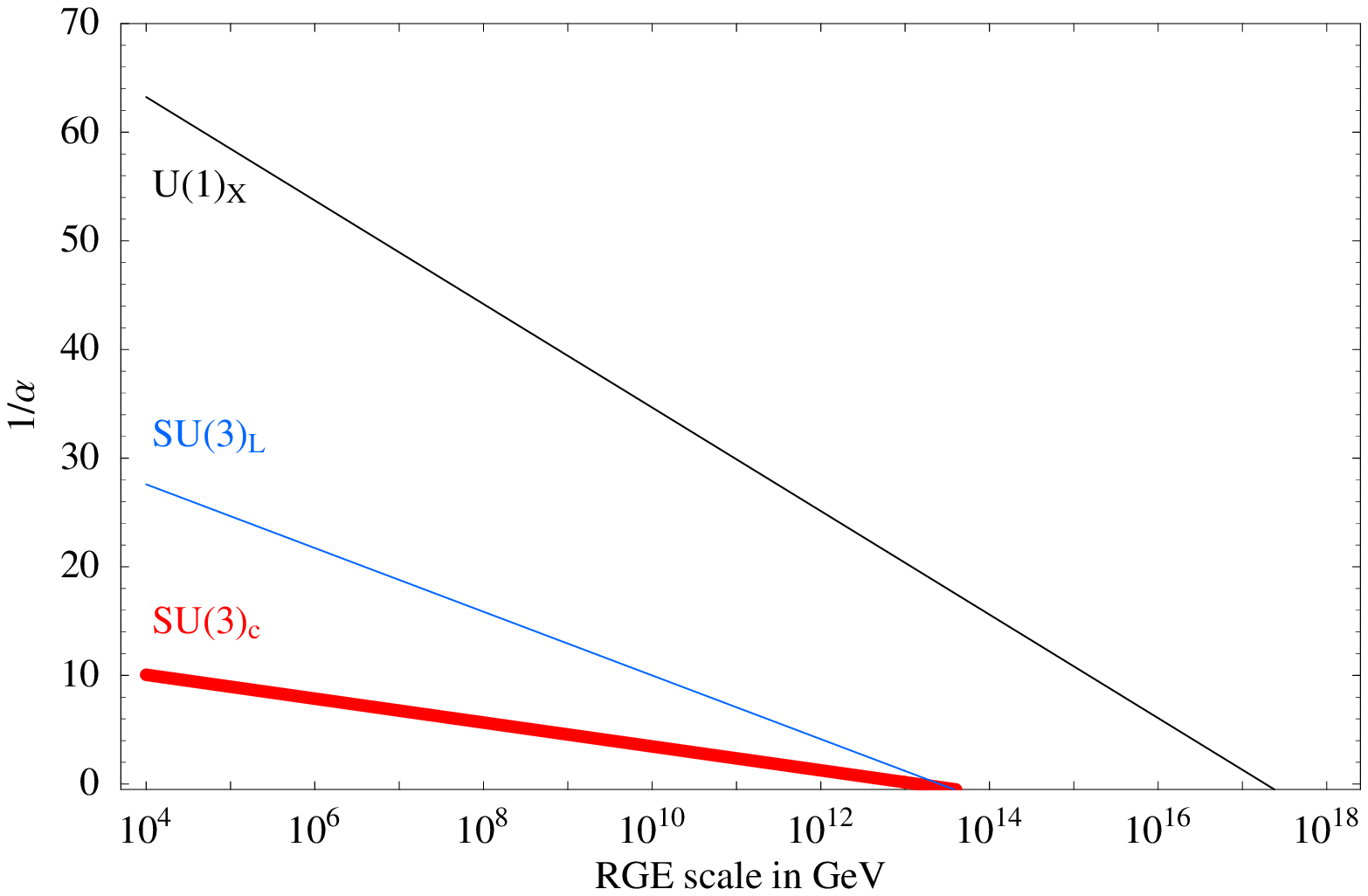}}
\end{center}\vspace{-8mm}
\caption{\label{fig:unification} \em One-loop running of the $\SU(3)_c\otimes\SU(3)_L\otimes {\rm U}(1)_X$
gauge couplings in a model with 3 light generations (a) without the matter fields added to generate the
top mass and with light colored Higgs triplets:
$\SU(6)$ unification can happen at a scale somewhat above the usual $\SU(5)$ GUT scale;
 (b) with the matter fields added to generate the top mass and Higgs
 and without light colored Higgs triplets:
 a dedicated non-perturbative analysis would be needed to test if unification happens.}
\end{figure}

With the matter content described above we can now calculate RG
evolution of the gauge couplings. The RGE $\beta$-function
coefficient above the SUSY-breaking and the little-Higgs scale are
$(1,4,10)$ for the $\SU(3)_c$, $\SU(3)_L$ and ${\rm U}(1)_X$ groups
respectively. We here assumed the SU(6)-unified normalization of the
$U(1)_X$ charges, in which the $H,\bar H$ fields have charges $\pm
1/\sqrt{6}$. We have not included the fields $Q\,\bar Q'$ introduced
for generating the top mass. One can see in
fig.~\ref{fig:unification}a that the gauge couplings can unify at a
scale somewhat higher than the usual SUSY SU(5) scale. Taking into
account the full $20$ of $\SU(6)$ does not affect unification, since
adding a complete $\SU(6)$ multiplet does not change the
differential running nor the unification scale. (To generate the top
Yukawa we only needed the 18 $Q',\bar{Q}'$ fields; the 20 contains two
extra singlets with $X$-charges $\pm1$). Fig.~\ref{fig:unification}b
shows the running in the minimal model, with a light 20 and without
light Higgs colored triplets: the QCD coupling hits a Landau pole at
energies around $10^{13}$ GeV.\footnote{The minimal model gives
successful perturbative SU(6) unification if SUSY is split, that is
if only the Higgses have light superpartners. We ignore this
possibility because it does not explain the hierarchy problem,
unless one resorts to antrophic arguments~\cite{Barr}.
Alternatively, perturbative $\SU(6)^2$ unification can be achieved
if duplicated copies of the $\SU(3)_c, \SU(3)_L, {\rm U}(1)_X$ gauge
groups survive unbroken down to low enough energies. Here we prefer
to insist on minimality.} However, this does not automatically imply
that there is something wrong. Techniques developed in the past
years allow us to analyze strongly interacting supersymmetric gauge
theories. The appearance of the QCD Landau pole implies that the
description valid at low energies breaks down at these
 intermediate energies.
At higher energies the theory can be described by a dual
weakly-coupled gauge group~\cite{Seiberg}. From a high-energy point
of view, the QCD  gluons are composite particles, which appear only
at low energies as the effective description of some more
fundamental gauge theory. The possibility that the SM gauge bosons
are composites has been discussed in~\cite{Seiberg}. In our case,
the gauge group above the QCD Landau pole would be $\SU(10)\otimes
\SU(3)_L\otimes {\rm U}(1)_X$. As it usually happens with product
gauge groups, now the $\SU(10)$ is asymptotically free. At slightly
higher energies the $\SU(3)_L$ gauge coupling hits its own Landau
pole. Thus we obtain a sort of a duality cascade, that is
reminiscent of the Klebanov-Strassler model~\cite{KS}. There it has
been conjectured that the duality cascades at high energies would
correspond to a string model on a warped throat, with the
supersymmetric SM appearing only at low energies. In the MSSM all
gauge couplings remain perturbative up to the Planck scale, so this
conjectured cascading behavior seems not well motivated. On the
contrary, it can arise in the model at hand. It would be interesting
to find out if the apparent unification of the couplings at
non-perturbative values will leave any consequences behind for the
cascading theory.

\subsection{The extra fermions}
In our  model fermions have a generation-universal and anomaly-free assignment of gauge charges.
The basic difference with respect to~\cite{Pok,Schmaltz} is that
singlet leptons are extended to $\SU(3)_L$ triplets rather than to $\SU(3)_L$ singlets.

Let us discuss the phenomenology of the extra  fermions.
Each generation contains an extra vector-like pair of
right-handed down-type quarks with mass $\sim f$ (contained in $D,Q$),
an extra vector-like pair of left-handed leptons with mass $\sim f$ (contained in $E,L$),
and two light $\SU(2)_L$ singlets (contained in the two $L$),
that we will name  `sterile neutrinos'.

The heavy extra fermions decay into ordinary fermions
via $\SU(3)_L$ gauge interactions.
The sterile neutrinos have a more interesting phenomenology.
They are light because renormalizable couplings do not allow any ${\cal O}(f)$ mass term.
A similar phenomenon was observed in~\cite{Dias}.
Unlike ordinary sterile neutrinos,
the singlets can be produced at accelerators thanks to their $\SU(3)_L$ interactions.

In cosmology, these $g=12$ degrees of freedom of sterile neutrinos
remain in thermal equilibrium down to $T_{\rm dec}\approx \GeV\cdot (f/20\TeV)^{4/3}$.
At this temperature the number of SM degrees of freedom is uncertain,
and varies by about a factor 2 depending
on whether $T_{\rm dec}$ is above or below
QCD phase transition occurring at imprecisely known temperature $T\sim m_\pi$.
We assume $g_{\rm SM}(T_{\rm dec})\sim 50$, such that sterile states
have temperature $T_{\rm s} = (10.75/g_{\rm SM}(T_{\rm dec}))^{1/3} T_\nu $.
Until relativistic, sterile states provide an extra energy density that is
conventionally expressed in terms of   `extra number of neutrinos' as
$\Delta N_\nu\sim 1.5$.
This excess is still considered acceptable~\cite{sterile,Olive},
and can be tested by CMB and BBN experiments.


One can modify the model such that sterile states get ${\cal O}(f)$ masses
(thereby avoiding phenomenological problems or signatures),
by adding more singlets that can get Dirac masses with the steriles.
Such extra singlets are automatically present in the full SU(6) model.
However, analogously to the usual see-saw,
it is considered more plausible that extra singlets also have Majorana masses much larger than $f,w$.
Integrating them out results into $\SU(3)_L$-invariant
Majorana neutrino mass operators, that now involve both active
and sterile neutrinos:
 $(LH)^2 \sim (\nu_L v + \nu_{\rm s} f)^2$.
The presence of two  $L$ fields per generation prevents a univocal prediction
for the sterile neutrino masses in terms of active neutrino masses.
Nevertheless, a typical pattern emerges:
sterile neutrinos with masses $m_{\nu_{\rm s}}\sim (f/v)^2 m_\nu \sim 10\eV$ and
active/sterile mixings
$v/f\sim 0.03$.
These masses and mixings are dangerously close to  the border of the presently
excluded region~\cite{sterile},
and possibly lie in the range suggested by the `sterile' interpretation of the LSND anomaly~\cite{LSND}.

\subsection{Consequences for the LHC}
The model presented here predicts various sets of new particles,
 that could be observable at the LHC.

 First of all, there are the `usual' supersymmetric particles.
 The stops are expected to be around a few hundred GeV.
 Depending on the SUSY-breaking model, this typically implies
 that most sparticles lie in this same energy range.
 Usual $R$-parity conservation is
expected in this model, so the traditional search methods for supersymmetry
should be applicable.

Next,  at the little-Higgs scale $f\sim $ few TeV there should be
a well-defined set of  extra gauge bosons,  Higgses, fermions,
and in particular the vector-like tops.
All these new particles are accompanied by their own supersymmetric partners.

Concerning gauge bosons, we expect a heavy $\SU(2)_L$ doublet $W'$
and a neutral $Z'$ with masses in the few TeV range. The lower
bounds on these masses from electroweak precision data are
$M_{W'}\circa{>}1.5\TeV$ and $M_{Z'}\circa{>}1.7\TeV$. Searching for
$W',Z'$ in collisions of SM fermions is analogous to searching for
$W,Z$ in $e,\bar e$ collisions. The $Z'$ couples to pairs of SM
fermions, giving a resonant contribution to four-fermion processes:
thanks to the clean signature and to the relatively large cross
section, the LHC sensitivity is in the multi-TeV range. This issue
has been studied for typical $Z'$ bosons~\cite{Tao,Atlas}. The model
univocally predicts all $Z'$ couplings.

Discovering the $W'$ is more difficult, because it couples one SM
fermion to one exotic fermion (a heavy quark and lepton, or a light
sterile) such that $W'$s need to be  pair-produced, giving no
resonance and rates more limited by the center-of-mass energy
available at the LHC. Single production of $W'$ is also possible,
but only thanks to couplings between the $W'$ and two SM fermions
suppressed by $v^2/f^2$~\cite{Tao}.

The vector-like top quarks could have masses in the $2\div3$ TeV
range. The LHC reach for this kind of particles  is expected to be
around $2\div2.5$ TeV~\cite{Tao,Atlas}. Not being odd under
$R$-parity, heavy tops can be singly produced.

The $\SU(3)_L$ partners of the light fermions
are similarly expected to have multi-TeV masses.
The LHC reach for these extra fermions is expected to be around a few TeV.

\section{Conclusions}
Supersymmetry and little-Higgs are respectively characterized by
the scales $m_{\rm soft}$ and $f$.
Present experimental bounds
\beq\label{eq:last} m_{\rm soft}\circa{>} \hbox{few hundred GeV}\qquad\hbox{and}\qquad
f\circa{>}\hbox{few TeV}\eeq
imply that typical supersymmetric models and
typical little-Higgs models are fine-tuned,
but that supersymmetric little-Higgs models
can still naturally explain electroweak-symmetry breaking,  provided that
the Higgs mass parameter arises as
\beq \label{eq:postlast} m_h^2 \sim \frac{{\cal O}(g^2,\lambda_t^2)}{(4\pi)^2}
m_{\rm soft}^2 \ln \frac{f}{m_{\rm soft}}\eeq
such that the Higgs gets a super-little VEV $v$.
This is however not the generic outcome of supersymmetric little-Higgs models.
Rather, $D$-terms typically break at tree level the global symmetry
 such that the Goldstone Higgs mass is not suppressed by a loop factor,
and furthermore non-Goldstone Higgses can dominantly contribute
to $\SU(2)_L$ breaking.

Therefore the goal consists in finding a non-generic and possibly
simple supersymmetric little-Higgs model that achieves
Eq.\eq{postlast}. Doublet/triplet splitting models based on the
SU(6) unification group already contain almost all necessary
ingredients, and can operate at low energies $f\sim$ few TeV
provided that SU(6) breaks as
$$\SU(6)\stackrel{M_{\rm GUT}}{\longrightarrow}   \SU(3)_c\otimes\SU(3)_L\otimes{\rm U}(1)_X
\stackrel{f}{\longrightarrow} \SU(3)_c\otimes\SU(2)_L\otimes{\rm
U}(1)_Y \stackrel{v}{\longrightarrow}\SU(3)_c\otimes{\rm U}(1)_{\rm
em}$$ rather that following the conventional SU(5) route. $\SU(3)_L$
is broken independently to the SM gauge group by a pair of higgs
triplets $H,\bar H$ and by an adjoint $\Sigma$, that contains
another pair of higgs doublets.
The resulting pseudo-Goldstone is light enough provided that
the superpotential coupling $\bar H \Sigma H$ has a coefficient smaller
than $\sim 10^{-2}$. The  little-Higgs structure is analogous to the
`simplest' little-Higgs, with various advantages. The model has an
anomaly-free, generation-independent and unified fermion content and
a `collective' source for the top Yukawa coupling.
Furthermore, thanks to the adjoint, in a well defined range of parameter
space the $D$-term problem simply does not arise.

Electroweak  symmetry breaking is then triggered
by the negative one-loop  contribution of the top sector to the mass squared of the Goldstone.
Little-Higgs model building is needed to satisfy the Higgs mass bound,
$m_h > 115\GeV$, adding an order one quartic self
interaction among the Goldstones.
This was achieved using a collective version of the sliding singlet mechanism.

Gauge couplings can unify in the presence of appropriate simple
SU(6)-embed\-dable  light chiral superfields. However, in the full
model, gauge couplings hit a Landau pole below the unification
scale, leading to a cascading gauge theory that could directly unify
into string theory on a warped throat. The fermion sector contains
light sterile neutrinos with superweak $\SU(3)_L$ interactions and
masses $f/v$ times higher than neutrino masses. The LHC should be
able to detect (beyond the ordinary MSSM superpartners) at least the
$Z'$ and hopefully some of the color triplet top partners as well.

\paragraph{Acknowledgments}
We thank Z. Berezhiani, M. Cirelli, A. Falkowski, T. Gregoire, M. Perelstein, S. Pokorski,
R. Rattazzi and J. Terning for useful discussions. C.C., G.M., and Y.S. would also like to thank the Aspen Center for Physics, where part of this work was done, for its hospitality.
The research of C.C.
is supported in part by the DOE OJI grant DE-FG02-01ER41206 and in part
by the NSF grants PHY-0139738  and PHY-0098631.
The work of G.M. is supported
in part by Department of Energy grant DE-FG02-91ER40674.
The research of Y.S. is supported by the U.S. Department
of Energy under contract W-7405-ENG-36.
The research of A.S.\ is supported in part by the European Programme `The Quest For Unification', contract MRTN-CT-2004-503369.

\footnotesize
\begin{multicols}{2}

\end{multicols}

\end{document}